\documentclass[pra,aps,twocolumn,showpacs]{revtex4}

\usepackage{times,epsfig,amssymb,amsfonts,amsmath}

\newcommand{\ket}[1]{|#1\rangle}
\newcommand{\bra}[1]{\langle #1|}
\newcommand{\proj}[1]{\ket{#1}\bra{#1}}

\begin{document}

\title{Entanglement evolution in multipartite cavity-reservoir
systems under local unitary operations}

\author{Wei Wen$^1$}
\author{Yan-Kui Bai$^{2,3}$}\email{ykbai@semi.ac.cn}
\author{Heng Fan$^3$}

\affiliation{$^1$ State Key Laboratory for Superlattices and
Microstructures, Institute of Semiconductors, Chinese Academy of
Sciences, P. O. Box 912, Beijing 100083, China\\
$^2$ College of Physical Science and Information Engineering and
Hebei Advance Thin Films Laboratory, Hebei Normal University,
Shijiazhuang, Hebei 050016, China\\
$^3$ Beijing National Laboratory for Condensed Matter Physics,
Institute of Physics, Chinese Academy of Sciences, Beijing 100080,
China}

\begin{abstract}
We analyze the entanglement evolution of two cavity photons being
affected by the dissipation of two individual reservoirs. Under an
arbitrary local unitary operation on the initial state, it is shown
that there is only one parameter which changes the entanglement
dynamics. For the bipartite subsystems, we show that the
entanglement of the cavity photons is correlated with that of the
reservoirs, although the local operation can delay the time at which
the photon entanglement disappears and advance the time at which the
reservoir entanglement appears. Furthermore, via a new defined
four-qubit entanglement measure and two three-qubit entanglement
measures, we study the multipartite entanglement evolution in the
composite system, which allows us to analyze quantitatively both
bipartite and multipartite entanglement
within a unified framework. In addition, we also discuss
the entanglement evolution with an arbitrary initial state.
\end{abstract}

\pacs{03.65.Ud, 03.65.Yz, 03.67.Mn}

\maketitle

\section{Introduction}

As one of the most subtle phenomena in many-body systems, quantum
entanglement has now been an important physical resource widely used
in quantum communication and quantum computation
\cite{hor09rev,ple07qic}. Therefore it is fundamental to
characterize the entanglement nature in quantum systems, especially
at a quantitative level. Till now, although bipartite entanglement
is well understood in many aspects, the entanglement in multipartite
systems is far from clear and thus deserve further exploration.

Entanglement dynamical behavior is an important property in
practical quantum information processing. This is because
entanglement is fragile and always decays due to unwanted
interactions between the system and its environment. A theoretical
study of two-atom spontaneous emission shows that entanglement does
not always decay in an asymptotic way and it can be corrupted in a
finite time \cite{tyu04prl}, which is referred to as entanglement
sudden death (ESD). Some earlier studies also pointed out this fact
that even a very weakly dissipative environment can disentangle the
quantum system in a finite time
\cite{zyc01pra,akr01pra,daf03pra,dod04pra}. The ESD phenomenon has
recently received a lot of attentions
\cite{ban06jpa,san06pra,der06pra,sun07pra,lfw09pra,fan09pra,ajh10pra,ysw10pra}
(see also a review paper \cite{tyu09sci} and references therein),
and, experimentally, it has been detected in photon \cite{alm07sci}
and atom systems \cite{lau07prl}.

A deep understanding on the ESD phenomenon concerns the problem
where the lost entanglement goes. To answer the question, it is proper to
enlarge the system to include its environment. Recently, L\'opez
\emph{et al} analyzed the entanglement evolution in a composite
system consisting of entangled cavity photons with individual
reservoirs \cite{clo08prl}, and show that the entanglement sudden
birth (ESB) of reservoir-reservoir subsystem must happen whenever
the ESD of cavity-cavity subsystem occurs. Moreover, in Ref. \cite{byw09pra},
Bai \emph{et al} presented a entanglement monogamy relation in
multipartite systems and analyzed quantitatively the bipartite
entanglement transfer in the multipartite cavity-reservoir system.

However, in the above analysis, the multipartite entanglement in the
composite cavity-reservoir system is not well characterized,
although the residual entanglement \cite{byw09pra} can indicate its
existence. Moreover, the authors only consider the symmetric initial
state like $\ket{\phi}=\alpha\ket{00}+\beta\ket{11}$. When the initial state
is asymmetric, the entanglement evolution can be very different. For
example, a $\sigma_{x}$ operation acting on the symmetric state can
change the evolution of the entangled cavity photons from the ESD
route to the asymptotic decay route, although the two kinds of
initial states have the equal entanglement. Therefore, it is
desirable to consider the entanglement dynamical behavior for
the asymmetric case and, particularly, find a good
entanglement measure to characterize the genuine multipartite
entanglement evolution.

In this paper, for the asymmetric initial state modulated by an
arbitrary local unitary (LU) operation, we analyze its entanglement
evolution in the multipartite cavity-reservoir system. In Sec. II,
we derive the effective output state under the LU operation, in
which there is only one parameter affecting the entanglement dynamics.
In Sec. III, we analyze the bipartite entanglement transfer
in the composite system, and point out the cavity photon entanglement
is still correlated with the reservoir entanglement although
the local operation can delay the ESD time and advance the ESB time.
In Sec. IV, the multipartite entanglement evolution is studied via a new
defined four-qubit entanglement measure and two three-qubit
entanglement measures. In Sec. V, within a unified framework, we
investigate the relation between bipartite entanglement transfer and
multipartite entanglement transition in the composite system.
Finally, we discuss the entanglement evolution with an arbitrary
initial state and give a brief conclusion in Sec. VI.

\section{The effective output state under the LU operation}

Before the derivation of the effective output state under the LU
operation, we first recall the multipartite cavity-reservoir system.
In Ref. \cite{clo08prl}, L\'opez \emph{et al} considered two
entangled cavity photons being affected by the dissipation of two
individual $N$-mode reservoirs where the interaction of a single
cavity-reservoir system is described by the Hamiltonian
\begin{equation}\label{1}
\hat{H}=\hbar \omega
\hat{a}^{\dagger}\hat{a}+\hbar\sum_{k=1}^{N}\omega_{k}
\hat{b}_k^{\dagger}\hat{b}_k+\hbar\sum_{k=1}^{N}g_{k}(\hat{a}
\hat{b}_{k}^{\dagger}+\hat{b}_{k}\hat{a}^{\dagger}).
\end{equation}
The authors analyzed the entanglement evolution with the symmetric
initial state
\begin{equation}\label{2}
\ket{\Phi_{0}}=(\alpha\ket{00}+\beta\ket{11})_{c_1c_2}\ket{00}_{r_1r_2},
\end{equation}
in which the reservoirs are in the vacuum state and the quantum
state of cavity photons is invariant under the permutation of the
qubits $c_1$ and $c_2$. They show that, along the time evolution,
the ESD of two photons can happen when the initial state amplitudes
satisfy the condition $\alpha<\beta$, and this procedure is \emph{necessarily}
related to the ESB of two reservoirs.

Now, we consider the asymmetric initial state modulated by an
arbitrary single-qubit LU operation. Without loss of generality, we
assume that the operation acts on the first cavity, and then the
initial state can be written as
\begin{equation}\label{3}
\ket{\Phi_{0}^a}=U_{c_1}\ket{\Phi_{0}}.
\end{equation}
For an arbitrary single qubit LU operation, one can decompose it as
\cite{nie00book}
\begin{equation}\label{4}
U(\zeta,\eta,\gamma,
\delta)=e^{i\zeta}R_{z}(\eta)R_{y}(\gamma)R_{z}(\delta),
\end{equation}
where the $e^{i\zeta}$ is a global phase shift and
$R_{k}(\theta)=\mbox{exp}(-i\theta \sigma_k/2)$ is the rotation
along the $k(=y,z)$ axis with the $\sigma_k$ being the Pauli matrix.
In this case, the output state under the time evolution is
\begin{eqnarray}\label{5}
\ket{\Phi_t}&=&U_{c_1r_1}(\hat{H},t)\otimes
U_{c_2r_2}(\hat{H},t)\ket{\Phi_0^a}\nonumber\\
&\simeq&U_{c_1r_1}(\hat{H}^{\prime},t) \otimes
U_{c_2r_2}(\hat{H}^{\prime\prime},t)[R_y(\gamma)_{c_1}\ket{\Phi_0}],
\end{eqnarray}
where
$\hat{H}^{\prime}=R_z^{\dagger}(\eta)_{c_1}\hat{H}R_z(\eta)_{c_1}$,
$\hat{H}^{\prime\prime}=R_z^{\dagger}(\delta)_{c_2}\hat{H}R_z(\delta)_{c_2}$,
and the $\simeq$ means the states on two sides are equivalent up to
some LU operations (for a detail derivation, see the
appendix). After considering the effect of the evolution
$U_{c_1r_1}(\hat{H}^{\prime},t)$ on the entanglement dynamics,
we find that it is equivalent to that of the evolution
$U_{c_1r_1}(\hat{H},t)$ (in the appendix, we give the proof).
The case for the evolution $U_{c_2r_2}(\hat{H}^{\prime\prime},t)$
is similar. Then Eq. (5) can be rewritten as
\begin{equation}\label{6}
\ket{\Phi_t} \simeq U_{c_1r_1}(\hat{H},t)\otimes U_{c_2r_2}(\hat{H},t)
[R_y(\gamma)_{c_1}\ket{\Phi_0}],
\end{equation}
which means that, under an arbitrary LU operation
$U_{c_1}(\zeta,\eta,\gamma,\delta)$, the entanglement evolution is
only sensitive to the rotation $R_y(\gamma)_{c_1}$.

Therefore, the effective initial state for the entanglement
evolution is
\begin{equation}\label{7}
\ket{\Psi_0}=R_y(\gamma)_{c_1}\ket{\Phi_0}=(\alpha\ket{\tilde{0}0}
+\beta\ket{\tilde{1}1})_{c_1c_2}\ket{00}_{r_1r_2},
\end{equation}
in which the new basic vectors are
$\ket{\tilde{0}}=\mbox{cos}(\gamma/2)\ket{0}+\mbox{sin}(\gamma/2)\ket{1}$
and
$\ket{\tilde{1}}=-\mbox{sin}(\gamma/2)\ket{0}+\mbox{cos}(\gamma/2)\ket{1}$.
For the output state, we use the approximation \cite{clo08prl}
\begin{equation}\label{8}
U(\hat{H},t)_{cr}\ket{10}=\xi\ket{10}+\chi\ket{01},
\end{equation}
where the amplitudes are $\xi(t)=\mbox{exp}(-\kappa t/2)$ and
$\chi(t)=[1-\mbox{exp}(-\kappa t)]^{1/2}$ in the limit of
$N\rightarrow \infty$ for a reservoir with a flat spectrum. Then the
effective output state has the form
\begin{eqnarray}\label{9}
\ket{\Psi_t}&=&\alpha(\mbox{cos}\frac{\gamma}{2}\ket{00}
+\mbox{sin}\frac{\gamma}{2}\ket{\phi_t})_{c_1r_1}\ket{00}_{c_2r_2}\nonumber\\
&&-\beta(\mbox{sin}\frac{\gamma}{2}\ket{00}
-\mbox{cos}\frac{\gamma}{2}\ket{\phi_t})_{c_1r_1}\ket{\phi_t}_{c_2r_2},
\end{eqnarray}
where $\ket{\phi_t}=\xi(t)\ket{10}_{cr}+\chi(t)\ket{01}_{cr}$ and
the parameter $\gamma$ being chosen in the range
$[0,\pi]$.

\section{Two-qubit entanglement evolution under the LU operation}

According to the effective output state $\ket{\Psi_t}$ in Eq. (9),
we can derive the density matrices of different subsystems and analyze
their entanglement dynamical behaviors.
We first consider the subsystem of two cavity photons, for which its
density matrix is
\begin{equation}\label{10}
\rho_{c_1c_2}(t)=\psi_1+\psi_2+\psi_3+\psi_4,
\end{equation}
where $\psi_i=\proj{\psi_i}$ and the four non-normalized pure state
components are
$\ket{\psi_1}=\alpha\mbox{cos}(\gamma/2)\ket{00}+\alpha\mbox{sin}(\gamma/2)
\xi\ket{10}-\beta\mbox{sin}(\gamma/2)\xi\ket{01}+
\beta\mbox{cos}(\gamma/2)\xi^2\ket{11}$,
$\ket{\psi_2}=\beta\mbox{sin}(\gamma/2)\chi\ket{00}-\beta\mbox{cos}(\gamma/2)
\xi\chi\ket{10}$,
$\ket{\psi_3}=\alpha\mbox{sin}(\gamma/2)\chi\ket{00}+\beta\mbox{cos}(\gamma/2)
\xi\chi\ket{01}$, and
$\ket{\psi_4}=\beta\mbox{cos}(\gamma/2)\chi^2\ket{00}$, respectively.
For the two reservoirs, its density matrix is similar to that of the
cavity photons and the following relation holds
\begin{equation}\label{11}
\rho_{r_1r_2}(t)=S_{\xi\leftrightarrow\chi}[\rho_{c_1c_2}(t)],
\end{equation}
where $S_{\xi\leftrightarrow\chi}$ exchanges the
parameters $\xi$ and $\chi$ (\emph{i.e.}, $\xi\rightarrow\chi$ and
$\chi\rightarrow\xi$).

Based on the previous analysis in Ref. \cite{byw09pra},
we choose the square of the concurrence to
characterize the two-qubit entanglement evolution. The concurrence is
defined as \cite{woo98prl} $C(\rho_{ij})=\mbox{max}(0,
\sqrt{\lambda_1}-\sqrt{\lambda_2}-\sqrt{\lambda_3}-\sqrt{\lambda_4})$
with the decreasing nonnegative real numbers $\lambda_{i}$ being the
eigenvalues of the matrix
$R_{ij}=\rho_{ij}(\sigma_y\otimes\sigma_y)\rho_{ij}^{\ast}
(\sigma_y\otimes\sigma_y)$. After computing the eigenvalues of the matrices
$R_{c_1c_2}$ and $R_{r_1r_2}$ \cite{expl2}, we can obtain
\begin{eqnarray}\label{12}
C_{c_1c_2}^2(t)&=&4[\mbox{max}(|\alpha\beta\xi^2|-|\beta\xi\chi|^2\mbox{cos}^2
(\gamma/2),0)]^2,\nonumber\\
C_{r_1r_2}^2(t)&=&4[\mbox{max}(|\alpha\beta\chi^2|-|\beta\xi\chi|^2\mbox{cos}^2
(\gamma/2),0)]^2.
\end{eqnarray}
Combining the concurrences and the expressions of $\xi(t)$ and
$\chi(t)$ in Eq. (8), we know that, for a given asymmetric initial state
(given the parameters $\alpha$, $\beta$ and $\gamma$), the cavity photons entanglement
decreases and the reservoir entanglement increases along with the time
evolution.

It is still an unsolved problem whether or not the ESD of cavity
photons and the ESB of the reservoirs are correlated in the asymmetric
case. With the conditions $C_{c_1c_2}(t)=0$ and $C_{r_1r_2}(t)=0$,
we can deduce the times of the ESD and the ESB, which have the forms
\begin{eqnarray}\label{13}
t_{ESD}(\rho_{c_1c_2})&=&-\frac{1}{\kappa}\mbox{ln}
\left(1-\frac{\alpha}{\beta\cdot\mbox{cos}^2
(\gamma/2)}\right),\nonumber\\
t_{ESB}(\rho_{r_1r_2})&=&\frac{1}{\kappa}\mbox{ln}
\frac{\beta\cdot\mbox{cos}^2(\gamma/2)}{\alpha},
\end{eqnarray}
where the parameter $\kappa$ is the dissipative constant (note that
$\xi=\mbox{exp}(-\kappa t/2)$ in the entanglement evolution).
According to the two times, we can derive that the ESD of two photons
occurs when $\beta\cdot\mbox{cos}^2(\gamma/2)>\alpha$, as is the case for the
ESB of two reservoirs. This means that the correlation between
the ESD and the ESB \emph{still} holds for the asymmetric initial
states, i.e., the ESB of the reservoirs must happen when
the ESD of cavity photons occurs.

\begin{figure}
\begin{center}
\epsfig{figure=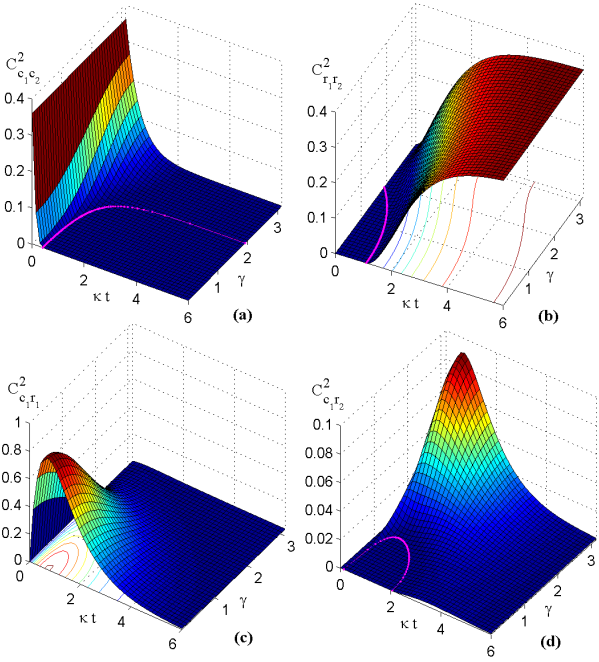,width=0.45\textwidth}
\end{center}
\caption{(Color online) Different two-qubit entanglement as a function
of the time evolution $\kappa t$ and the rotation parameter $\gamma$.}
\end{figure}

As an example, we choose the initial state parameters as
$\alpha=1/\sqrt{10}$ and $\beta=3/\sqrt{10}$. In Fig. 1(a),
the concurrence $C_{c_1c_2}^2$ is plotted as a function of the time
$\kappa t$ and the rotation parameter $\gamma$. For a fixed value of
$\kappa t$, the photon entanglement increases with the parameter $\gamma$.
When the $\gamma$ is given, the $C_{c_1c_2}^2$ decreases along the time
$\kappa t$. The ESD line (the purple line) is also plotted in the
figure, where the $\gamma$ can delay the ESD time. It is
interesting that the entanglement evolution changes to the
asymptotical decay route before the $\gamma$ attains to the value
$\pi$, and the critical value is
$\gamma=2\mbox{arccos}\sqrt{1/3}\approx 1.91063$. In Fig. 1(b), the
entanglement evolution of $C_{r_1r_2}^2$ is plotted, where the
parameter $\gamma$ can increase the reservoir entanglement and
advance the ESB time (the purple line). The critical value for the
route transition is also $\gamma=2\mbox{arccos}\sqrt{1/3}$.
Moreover, depending on the value of the $\gamma$, the ESB
can manifest before, simultaneously and after the ESD.

The two-qubit entanglement of subsystem $c_1r_1$ has the form
\begin{equation}\label{14}
C_{c_1r_1}^2(t)=\xi^2\chi^2[1+(\beta^2-\alpha^2)\mbox{cos}(\gamma)]^2.
\end{equation}
In Fig. 1(c), the concurrence is plotted as a function of the parameters
$\kappa t$ and $\gamma$. The maximum of $C_{c_1r_1}^2(t)$ appears
at the time $\kappa t=\mbox{ln}2$, and the entanglement decreases
with the $\gamma$. However, the $\gamma$
does not change the entanglement of subsystem $c_2r_2$, because
the rotation $R_y(\gamma)$ acts on the first cavity. For the
subsystems $c_1r_2$ and $c_2r_1$, we can get that they have the
equal entanglement, which can be expressed as
\begin{equation}\label{15}
C_{c_1r_2}^2(t)=4[\mbox{max}(|\alpha\beta\xi\chi|-|\beta\xi\chi|^2
\mbox{cos}^2(\gamma/2),0)]^2.
\end{equation}
In Fig. 1(d), the concurrence is plotted.
When $\gamma=0$, the entanglement evolution experiences the
ESD at the time $\kappa t=\mbox{ln}[3(3-\sqrt{5})/2]$ and the ESB at
the time $\kappa t=\mbox{ln}[3(3+\sqrt{5})/2]$ (the two
intersections between the purple line and the $\kappa t$ axis), then
the entanglement changes asymptotically. Along with the increase of
the parameter $\gamma$, the time window between the ESD and ESB
decreases, and the window become a point when
$\gamma=2\mbox{arccos}\sqrt{2/3}\approx 1.23096$. After this value,
both the ESD and the ESB phenomena disappear.

\section{Multipartite entanglement evolution under the LU operation}
Before analyzing the entanglement evolution, we first consider how
to characterize the multipartite entanglement in the composite system.
In Ref. \cite{clo08prl}, the multipartite concurrence $C_N$ \cite{car04prl}
can not characterize completely the genuine multipartite
entanglement, due to its nonzero value for two
Bell states. In Ref. \cite{byw09pra}, it is shown that the genuine
multipartite entanglement can be indicated by the two-qubit residual
entanglement
\begin{equation}\label{16}
M_{c_1r_1}(\Phi_t)=C_{c_1r_1|c_2r_2}^2(t)-\sum
C_{i^{\prime}j^{\prime}}^2(t),
\end{equation}
where the sum subscripts $i^{\prime}\in\{c_1, r_1\}$ and
$j^{\prime}\in\{c_2, r_2\}$, respectively. However, the entanglement
monotone property of $M_{c_1r_1}$ is not clear, even for the case of
symmetric initial states.

The average multipartite entanglement may quantify the genuine
multiqubit entanglement based on much numerical analysis, which is
defined as \cite{byw07pra}
\begin{eqnarray}\label{17}
E_{ms}(\Psi_4)=\frac{\sum_i \tau_i(\rho_i)-2\sum_{i>j}
C_{ij}^2(\rho_{ij})}{4},
\end{eqnarray}
where the $\tau_i=2(1-\mbox{tr}\rho_i^2)$ is the linear entropy and
the $C_{ij}$ is the concurrence. For the four-qubit cluster-class
states, an analytical proof of the entanglement monotone property
for the $E_{ms}$ was given in Refs. \cite{baw08pra,ren08pra}.

For the effective output state $\ket{\Psi_t}$ in Eq. (9), we can compute
the average multipartite entanglement $E_{ms}$ and the residual entanglement
$M_{c_1r_1}$. After comparing the two measures, we can obtain that
they are equivalent up to a constant factor $2$. Therefore,
we can define entanglement measure
\begin{equation}\label{18}
E_{BB}(\Psi_t)=M_{c_1r_1}(\Psi_t)=2E_{ms}(\Psi_t),
\end{equation}
which quantifies the genuine multipartite entanglement between the
blocks $c_1r_1$ and $c_2r_2$ and its entanglement monotone property
is based on the numerical analysis on the average multipartite
entanglement $E_{ms}$. In Fig. 2, the $E_{BB}$ is plotted as a function of
the parameters $\kappa t$ and $\gamma$, where the initial
state parameters are chosen as $\alpha=1/\sqrt{10}$ and $\beta=3/\sqrt{10}$.
When $\gamma=0$, the $E_{BB}(\kappa t)$ increases from $0$ to $0.36$
in the region $\kappa t\in\{0,\mbox{ln}(3/2)\}$, then it keeps invariant
until the time $\kappa t=\mbox{ln}3$, finally, the $E_{BB}(\kappa t)$ decreases
asymptotically. With the increase of the $\gamma$, the width of the
plateau decreases and can be expressed as
\begin{equation}\label{19}
t_w=\mbox{ln}[3\mbox{cos}^2(\gamma/2)-1].
\end{equation}
When $\gamma=2\mbox{arccos}(\sqrt{2/3})$, the width changes
to zero and the evolution time is $\kappa t=\mbox{ln}2$. After this
value, the block-block entanglement decreases along with the $\gamma$,
and vanishes when $\gamma=\pi$.

\begin{figure}
\begin{center}
\epsfig{figure=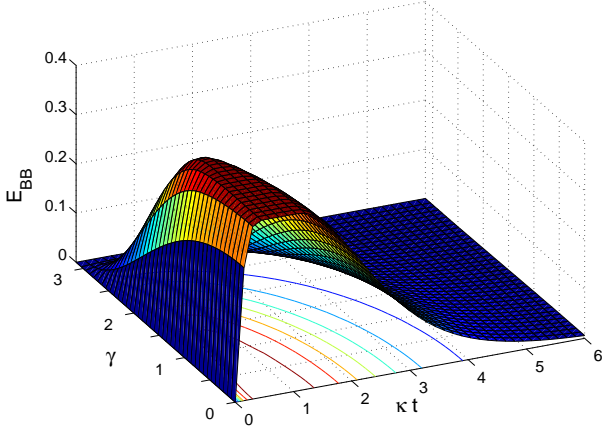,width=0.45\textwidth} \caption{(Color
online) Block-block entanglement as a function of the parameters
$\kappa t$ and $\gamma$.}
\end{center}
\end{figure}

The genuine tripartite entanglement in the composite system can
be quantified by the mixed state three-tangle \cite{won01pra}
\begin{equation}\label{20}
\tau_3(\rho_{ijk})=\mbox{min}\sum_{\{p_x,\varphi^{(x)}_{ijk}\}}
p_x\tau(\varphi_{ijk}^{(x)}),
\end{equation}
where $\tau(\varphi^{(x)}_{ijk})=\tau_i-C_{ij}^2-C_{ik}^2$
\cite{ckw00pra} is the pure state three-tangle and the minimum runs
over all the pure state decompositions of $\rho_{ijk}$.
The reduced density matrix of subsystem $c_1r_1c_2$ can be
written as
\begin{equation}\label{21}
\rho_{c_1r_1c_2}(t)=\varphi_1(t)+\varphi_2(t),
\end{equation}
where the non-normalized pure state components are
$\ket{\varphi_1(t)}=\alpha\mbox{cos}(\gamma/2)\ket{000}-\beta\mbox{sin}(\gamma/2)
\xi\ket{001}+\alpha\mbox{sin}(\gamma/2)\chi\ket{010}+\alpha\mbox{sin}(\gamma/2)
\xi\ket{100}+\beta\mbox{cos}(\gamma/2)\xi^2\ket{101}+\beta\mbox{cos}\xi\chi\ket{011}$
and
$\ket{\varphi_2(t)}=\beta\mbox{sin}(\gamma/2)\chi\ket{000}-\beta\mbox{cos}
(\gamma/2)\chi^2\ket{010}-\beta\mbox{cos}(\gamma/2)\xi\chi\ket{100}$,
respectively. It is obvious that the $\ket{\varphi_2}$ is a
separable state and its three-tangle is zero. Moreover, for the component
$\ket{\varphi_1}$, we can derive $\tau(\varphi_1)=0$.
So, the decomposition in Eq. (21) is the optimal and the mixed
state three-tangle $\tau_3(\rho_{c_1r_1c_2})$ is zero.
Similarly, we can obtain that all the other mixed state
three-tangles $\tau_3(\rho_{ijk})$ are zero.

Although all the $\tau_3(\rho_{ijk})$ are zero in the entanglement
evolution, the three-qubit states are still entangled in the
qubit-block form \cite{byw08pra,loh06prl}, which is not equivalent
to the mixed state three-tangle and can not be accounted for the
two-qubit entanglement. The qubit-block entanglement characterizes
the genuine three-qubit entanglement under bipartite cut between a qubit
and a block of qubits, and can be defined as
$E_{q-B}(\rho_{i|jk})=C_{i|jk}^2-C_{ij}^2-C_{ik}^2$,
in which the $C_{i|jk}$ quantifies bipartite entanglement
between the qubits $i$ and $jk$.
For the subsystems $c_1c_2r_2$ and $r_1c_2r_2$, their qubit-block
entanglement are
\begin{eqnarray}\label{22}
E_{q-B}(\rho_{c_1|c_2r_2})&=&C_{c_1|c_2r_2}^2-C_{c_1c_2}^2-C_{c_1r_2}^2
\nonumber\\
E_{q-B}(\rho_{r_1|c_2r_2})&=&C_{r_1|c_2r_2}^2-C_{r_1r_2}^2-C_{c_2r_1}^2,
\end{eqnarray}
where  $C_{c_1|c_2r_2}^2=4\alpha^2\beta^2\xi^2$,
$C_{r_1|c_2r_2}^2=4\alpha^2\beta^2\chi^2$, and the expressions of two-qubit
concurrences $C_{ij}^2$ are given in Eqs. (12), (14) and (15).

\begin{figure}
\begin{center}
\epsfig{figure=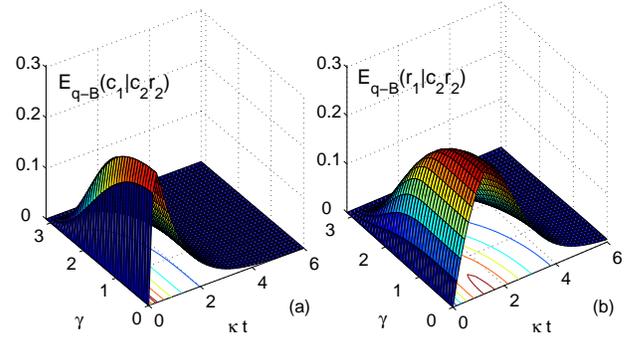,width=0.45\textwidth} \caption{(Color
online) (a) Qubit-block entanglement $E_{q-b}(c_1|c_2r_2)$ and
(b) qubit-block entanglement $E_{q-b}(r_1|c_2r_2)$ as a function
of the parameters $\kappa t$ and $\gamma$.}
\end{center}
\end{figure}

In Fig. 3, we plot the qubit-block entanglement as a function of the
parameters $\kappa t$ and $\gamma$, where the initial state parameters
are chosen as $\alpha=1/\sqrt{10}$
and $\beta=3/\sqrt{10}$. For a given value of the $\gamma$, the
qubit-block entanglement $E_{c_1|c_2r_2}$ (in Fig. 3(a)) increases
first with the time $\kappa t$, and then decreases with the $\kappa
t$ after attaining to its maximal value. Along with the increase of
the $\gamma$, the maximal value of $E_{c_1|c_2r_2}$ decreases. For
the qubit-block entanglement $E_{r_1|c_2r_2}$ (in Fig. 3(b)), the
trend of entanglement evolution is similar.
In Refs. \cite{byw08pra,byw09pra}, it is pointed
out that the qubit-block entanglement comes from the genuine multipartite
entanglement in the enlarged pure state system. Here, for the
multipartite cavity-reservoir system, we can derive the following relation
\begin{equation}\label{23}
E_{BB}(\Psi_t)=E_{q-B}(\rho_{c_1|c_2r_2}(t))
+E_{q-B}(\rho_{r_1|c_2r_2}(t)),
\end{equation}
which means that the qubit-block entanglement comes from the genuine
block-block entanglement in the composite system.

\section{Entanglement transfer and entanglement
transition under the LU operation}

Because the evolution $U_{c_1r_1}(\hat{H},t)\otimes
U_{c_2r_2}(\hat{H},t)$ are two local unitary operations under the
partition $c_1r_1|c_2r_2$, the bipartite entanglement
$C_{c_1r_1|c_2r_2}^2$ is invariant, and the following relation holds
\begin{equation}\label{24}
C_{c_1r_1|c_2r_2}^2(\Psi_t)=E_{BB}(t)+\sum
C_{i^{\prime}j^{\prime}}^2(t)=4\alpha^2\beta^2,
\end{equation}
where $i^{\prime}\in\{c_1, r_1\}$ and
$j^{\prime}\in\{c_2, r_2\}$, respectively.
Therefore, in the multipartite cavity-reservoir system, we can
characterize the entanglement evolution under a unified framework, where
the two qubit entanglement transfer is quantified by the concurrence and
the multipartite entanglement transition is quantified by the
block-block entanglement $E_{BB}$. In Fig.4, the entanglement evolution
modulated by different value of $\gamma$ is plotted, where
$\gamma_1=2\mbox{arccos}(\sqrt{2/3})$, $\gamma_2=2\mbox{arccos}(\sqrt{1/3})$,
and the initial state parameters are $\alpha=1/\sqrt{10}$ and
$\beta=3/\sqrt{10}$.

\begin{figure}
\begin{center}
 \epsfig{figure=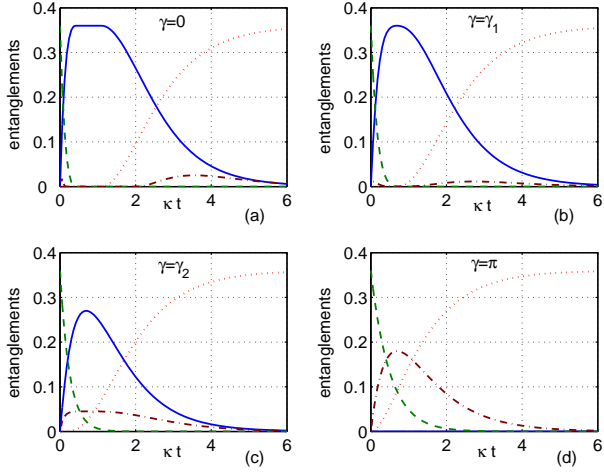,width=0.45\textwidth}
\caption{(Color online) Entanglement evolutions modulated by
different parameter values: (a) $\gamma=0$, (b) $\gamma=\gamma_1$,
(c) $\gamma=\gamma_2$ and (d) $\gamma=\pi$. The entanglements
$C_{c_1c_2}^2$, $C_{r_1r_2}^2$, $C_{c_1r_2}^2+C_{c_2r_1}^2$ and
$E_{BB}$ are represented by the green dashed line, red dotted line,
brown dot-dashed line and blue solid line, respectively.}
\end{center}
\end{figure}

In Fig. 4(a), the parameter is chosen as $\gamma=0$, which
corresponds to the symmetric initial state. In the time interval
$\kappa t\in(0,\mbox{ln}(3/2))$, a part of the initial photon-photon
entanglement $C_{c_1c_2}^2$ first transfers to the subsystems
$c_1r_2$ and $c_2r_1$ (the brown dot-dashed line where a factor $5$
is multiplied), then the remaining photon-photon entanglement and
the cavity-reservoir entanglement transition completely to the
genuine block-block entanglement $E_{BB}$ (the blue solid line).
Along with the time evolution, the block-block entanglement keeps
invariant and is immune to the cavity-reservoir interaction in the
time interval $\kappa t\in [\mbox{ln}(3/2),\mbox{ln}3]$. Finally,
when $\kappa t \in(\mbox{ln}3,6]$, the multipartite entanglement
$E_{BB}$ transitions to the two-qubit reservoir-reservoir
entanglement (the red dotted line) and the cavity-reservoir
entanglement. When the parameter $\gamma\in (0,\gamma_1]$, the
trends of bipartite and multipartite entanglement evolutions are
similar to those when $\gamma=0$, but the immune region of
the block-block entanglement decreases with the parameter and the
other evolution regions extend. In Fig. 4(b), the parameter is
chosen as $\gamma=\gamma_1$, where the plateau
region of the $E_{BB}$ changes to a point ($\kappa t=\mbox{ln}2$)
and the procedures of entanglement transfer and entanglement
transition need more time.

When the parameter $\gamma\in (\gamma_1,\gamma_2]$, the initial
photon entanglement transfers to not only the subsystems $c_1r_2$
and $c_2r_1$ but also the subsystem $r_1r_2$, and the two-qubit
entanglement can not transition completely to the genuine
block-block entanglement. As shown in Fig. 4(c), the entanglement
transfer and entanglement transition are plotted when
$\gamma=\gamma_2$. Along with the increase of
the $\gamma$, the decay of the photon entanglement slows
down and the transfer ratio of the two-qubit entanglement increases.
At the same time, the transition ratio of the block-block entanglement
decreases. In Fig. 4(d), the parameter is chosen as $\gamma=\pi$, in which the
transition between the two-qubit entanglement and the multipartite
entanglement disappears, and the entanglement evolution consists of
only two-qubit entanglement transfer.

It should be pointed out that, in the unified framework of
entanglement evolution, the entanglement monotone property of $E_{BB}$ is
based on the numerical analysis on the average multipartite
entanglement $E_{ms}$ \cite{byw07pra}. The analytic proof is still an
open problem.

\section{Discussion and conclusion}

In the more general case, an arbitrary initial state has the form
$\ket{\Psi_0^g}=(\alpha_1\ket{00}+\alpha_2\ket{01}+\alpha_3\ket{10}
+\alpha_4\ket{11})_{c_1c_2}\ket{00}_{r_1r_2}$, which corresponds to
two LU operations $U_{c_1}(\zeta_1,\eta_1,\gamma_1,\delta_1)\otimes
U_{c_2}(\zeta_2,\eta_2,\gamma_2,\delta_2)$ acting on the symmetric
initial state. In this case, the analytical characterization for the
entanglement evolution is not available so far. However, the
correlation between the ESD of cavity photons and the ESB of
reservoirs still holds. This is because we can deduce the relation
\begin{equation}\label{25}
\rho_{c_1c_2}^g(\xi,\chi)=S_{\xi\leftrightarrow\chi}
[\rho_{r_1r_2}^g(\xi,\chi)],
\end{equation}
where the evolution
$U_{cr}(\hat{H},t)\ket{10}=\xi\ket{10}+\chi\ket{01}$ is used. Based
on this relation, we can obtain that when the ESD of cavity photons
occurs at the time $t_{ESD}=t_0$, the ESB of reservoirs will
necessarily happen at the time
$t_{ESB}=-(1/\kappa)\mbox{ln}[1-\mbox{exp}(-\kappa t_0)]$. Moreover,
the entanglement evolution is restricted by the monogamy relation
\begin{eqnarray}\label{26}
C_{c_1r_1|c_2r_2}^2(\ket{\Psi_0^g})&\geq&
C_{c_1c_2}^2(t)+C_{r_1r_2}^2(t)+C_{c_1r_2}^2(t),\nonumber\\
&&+C_{c_2r_1}^2(t)
\end{eqnarray}
and the multipartite entanglement can be indicated by the
two-qubit residual entanglement $M_{c_1r_1}$
\cite{byw09pra,osb06prl}.

The entanglement evolution with the asymmetric initial state is worth to
consider for other physical systems, for example, the atoms systems
\cite{tyu04prl}, quantum dots and spin chains etc.
\cite{dlo98pra,ssl01pns,xgw06pra}. Moreover, the dissipative
entanglement evolution has close relation with the type of the noise
environment. Therefore, the non-Markovian environment, the
correlated noises, and some operator channels
\cite{bel07prl,nov08pra,kon08nat,wan10epr} are also worth to study
in future.

In conclusion, we have investigated the entanglement evolution of
multipartite cavity-reservoir systems with the asymmetric initial
state. It is shown that there is only one parameter in the LU
operation affecting the entanglement dynamics, which can delay the
ESD of the photons, advance the ESB of the reservoirs, change the
evolution route of bipartite entanglement, and suppress
the multipartite entanglement. However, the correlation
between the ESD and the ESB still holds. Furthermore, by defining
the block-block entanglement, we analyze the multipartite entanglement
evolution in the composite system, which allows us to study quantitatively
both the entanglement transfer and the entanglement transition within
a unified framework. Finally, the entanglement evolution with an arbitrary
initial state is discussed.

\section*{Acknowledgments}
The authors would like to thank Prof. Z. D. Wang for many useful
discussions and suggestions. This work was supported by the National
Basic Research Program of China (973 Program) grant Nos.
2009CB929300 and 2010CB922904. Y.K.B. was also supported by the fund
of Hebei Normal University and NSF-China Grant No. 10905016.

\section*{Appendix}

We first prove Eq. (5). The output state under the time evolution is
\begin{eqnarray}\label{27}
\ket{\Phi_t}&=&U_{c_1r_1}(\hat{H},t)\otimes U_{c_2r_2}(\hat{H},t)
\ket{\Phi_0^a}\nonumber\\
&=&U_{c_1}U_{c_1}^{\dagger}U_{c_1r_1}(\hat{H},t)\otimes U_{c_2r_2}(\hat{H},t)
[U_{c_1}\ket{\Phi_0}],
\end{eqnarray}
where, in the second equation, the identity operator $I=U_{c_1}U_{c_1}^{\dagger}$
is inserted. Substituting the $U_{c_1}$ with the expression
$R_{z}(\eta)R_{y}(\gamma)R_{z}(\delta)$ (we neglect the
global phase $e^{i\zeta}$), we can obtain
\begin{eqnarray}\label{28}
\ket{\Phi_t}&=&U_{l}R_{z}^{\dagger}(\eta)U_{c_1r_1}(\hat{H},t)R_{z}(\eta)_{c_1}
\otimes U_{c_2r_2}(\hat{H},t)\nonumber\\
&&[R_y(\gamma)_{c_1}R_z(\delta)_{c_1}\ket{\Phi_0}]\\
&\simeq & U_{c_1r_1}(\hat{H}^{\prime},t)\otimes U_{c_2r_2}(\hat{H},t)
[R_y(\gamma)_{c_1}R_z(\delta)_{c_1}\ket{\Phi_0}]\nonumber
\end{eqnarray}
where the symbol $\simeq$ means the quantum states on the two sides are
equivalent up to the local unitary operation $U_l=U_{c_1}R_z^{\dagger}
(\delta)_{c_1}R_y^{\dagger}(\gamma)_{c_1}$ (note that entanglement is
invariant under local unitary transformation), and we use the relation
$R_z^{\dagger}(\eta)_{c_1}U_{c_1r_1}(\hat{H},t)R_z(\eta)_{c_1}=
U_{c_1r_1}(\hat{H}^{\prime},t)$ with $\hat{H}^{\prime}=R_z^{\dagger}(\eta)_{c_1}
\hat{H}R_z(\eta)_{c_1}$. Due to the symmetric property
of initial state $\ket{\Phi_0}$, we have the relation $R_z(\delta)_{c_1}
\ket{\Phi_0}=R_z(\delta)_{c_2}\ket{\Phi_0}$. Then the output state can be
expressed further as
\begin{eqnarray}\label{29}
\ket{\Phi_t}&=&U_{c_1r_1}(\hat{H}^{\prime},t)\otimes R_z(\delta)_{c_2}
R_z^{\dagger}(\delta)_{c_2}U_{c_2r_2}(\hat{H},t)R_z(\delta)_{c_2}\nonumber\\
&&[R_y(\gamma)_{c_1}\ket{\Phi_0}]\nonumber\\
&\simeq&U_{c_1r_1}(\hat{H}^{\prime},t)\otimes U_{c_2r_2}(\hat{H}^{\prime\prime},t)
[R_y(\gamma)_{c_1}\ket{\Phi_0}],
\end{eqnarray}
where we insert the identity operator $R_z(\delta)_{c_2}
R_z^{\dagger}(\delta)_{c_2}=I$ in the first equation and use the relation
$\hat{H}^{\prime\prime}=R_z^{\dagger}(\delta)_{c_2}
\hat{H}R_z(\delta)_{c_2}$ in the second equation.

Next, we will prove the effects of $\hat{H}^{\prime}$
and $\hat{H}^{\prime\prime}$
are equivalent to that of $\hat{H}$ in the entanglement evolution. Because
the Hilbert space of subsystem $c_1r_1$ is spanned by
$\ket{00},\ket{01},\ket{10}$, the creation and annihilation operators are
\begin{equation}\label{30}
\hat{a}^{\dagger}=\left(
\begin{array}{cc}
0 & 0 \\
1 & 0 \\
\end{array}
\right)
~\mbox{and}~ \hat{a}=\left(
\begin{array}{cc}
      0 & 1 \\
      0 & 0 \\
\end{array}
\right),
\end{equation}
respectively. With these expressions, the rotation operator
$R_{z}(\eta)$ can be rewritten as
\begin{equation}\label{31}
R_z(\eta)=A\cdot \hat{a}^{\dagger}\hat{a}+B,
\end{equation}
where the coefficients
$A=\mbox{exp}(i\eta/2)-\mbox{exp}(-i\eta/2)$ and
$B=\mbox{exp}(-i\eta/2)$, respectively.
After substituting the expression of $R_{z}(\eta)$ into the Hamiltonian
$\hat{H}^{\prime}$, we can derive
\begin{eqnarray}\label{32}
\hat{H}^{\prime}&=&R_z^{\dagger}(\eta)_{c_1}\hat{H}R_z(\eta)_{c_1}\nonumber\\
&=&(A \cdot
\hat{a}^{\dagger}\hat{a}+B)^{\dagger}[\hbar\omega\hat{a}^{\dagger}\hat{a}
+\hbar\sum_{k=1}^{N}\omega_{k}
\hat{b}_k^{\dagger}\hat{b}_k\nonumber\\
&&+\hbar\sum_{k=1}^{N}g_{k}(\hat{a}
\hat{b}_{k}^{\dagger}+\hat{b}_{k}\hat{a}^{\dagger})](A \cdot
\hat{a}^{\dagger}\hat{a}+B)\nonumber\\
&=&\hbar\omega\hat{a}^{\dagger}\hat{a}
+\hbar\sum_{k=1}^{N}\omega_{k}
\hat{b}_k^{\dagger}\hat{b}_k+\hbar\sum_{k=1}^{N}g_{k}(\hat{a}
\hat{b}_{k}^{\dagger}\cdot
e^{i\eta}\nonumber\\
&&+\hat{b}_{k}\hat{a}^{\dagger}\cdot e^{-i\eta})\nonumber\\
&=&V_{r_1}^{\dagger}(\eta)\hat{H}V_{r_1}(\eta),
\end{eqnarray}
where $V_r(\eta)=\mbox{diag}\{1,\mbox{exp}(-i\eta)\}$, and we used the
relations $\hat{a}^{\dagger}\hat{a}\hat{a}^{\dagger}\hat{a}=
\hat{a}^{\dagger}\hat{a}=\hat{N}$ and
$\hat{a}^{\dagger}\hat{N}\hat{b}_k=0$ \cite{expl1}. Similarly,
for the Hamiltonian $\hat{H}^{\prime\prime}$, we can obtain
\begin{equation}\label{33}
\hat{H}^{\prime\prime}=R_z^{\dagger}(\delta)_{c_2}\hat{H}R_z(\delta)_{c_2}
=V_{r_2}^{\dagger}(\delta)\hat{H}V_{r_2}(\delta)
\end{equation}
with $V_{r_2}(\delta)=\mbox{diag}\{1,\mbox{exp}(-i\delta)\}$. Therefore, the output
state in Eq. (29) can be written as
\begin{eqnarray}\label{34}
\ket{\Phi_t}
&=&V_{r_1}^{\dagger}(\eta)V_{r_2}^{\dagger}(\delta)U_{c_1r_1}(\hat{H},t)
\otimes U_{c_2r_2}(\hat{H},t)\nonumber\\
&&[R_y(\gamma)_{c_1}V_{r_1}(\eta)V_{r_2}(\delta)\ket{\Phi_0}].
\end{eqnarray}
In the above equation, the local unitary operation
$V_{r_1}^{\dagger}(\eta)V_{r_2}^{\dagger}(\delta)$ does not change
the entanglement evolution. Moreover, due to the reservoirs being in
the vacuum state, we have $V_{r_1}(\eta)V_{r_2}
(\delta)\ket{\Phi_0}=\ket{\Phi_0}$. Therefore, the effective
output state in the entanglement evolution has the form
\begin{equation}\label{35}
\ket{\Psi_t}=U_{c_1r_1}(\hat{H},t)\otimes U_{c_2r_2}(\hat{H},t)
[R_y(\gamma)_{c_1}\ket{\Phi_0}].
\end{equation}
This means that, for the asymmetric initial state modulated by
an arbitrary LU operation $U_{c_1}(\zeta,\eta,\gamma,\delta)$, the
entanglement evolution is only sensitive to the rotation $R_y(\gamma)$.


\begin{thebibliography}{99}

\bibitem{hor09rev}  R. Horodecki P. Horodecki, M. Horodecki and K. Horodecki,
                    Rev. Mod. Phys. \textbf{81}, 865 (2009).
\bibitem{ple07qic}  M. B. Plenio and S. Virmani, Quantum Inf. Comput.
                    \textbf{7}, 1 (2007).
\bibitem{tyu04prl}  Ting Yu and J. H. Eberly, Phys. Rev. Lett. \textbf{93},
                    140404 (2004); \emph{ibid}. \textbf{97}, 140403 (2006).
\bibitem{zyc01pra}  K.  \.Zyczkowski, P. Horodecki, M. Horodecki, and
                    R. Horodecki, Phys. Rev. A \textbf{65}, 012101 (2001).
\bibitem{akr01pra}  A. K. Rajagopal and R. W. Rendell, Phys. Rev. A
                    \textbf{63}, 022116 (2001).
\bibitem{daf03pra}  S. Daffer, K. Wodkiewicz, and J. K. Mclver,
                    Phys. Rev. A \textbf{67}, 062312 (2003).
\bibitem{dod04pra}  P. J. Dodd and J. J. Halliwell, Phys. Rev. A \textbf{69},
                    052105 (2004).
\bibitem{ban06jpa}  M. Ban, J. Phys. A, \textbf{39}, 1927 (2006).
\bibitem{san06pra}  M. F. Santos, P. Milman, L. Davidovich, and N. Zagury,
                    Phys. Rev. A \textbf{73}, 040305 (2006).
\bibitem{der06pra}  L. Derkacz and L. Jak\'{o}bczyk, Phys. Rev. A \textbf{74},
                    032313 (2006).
\bibitem{sun07pra}  Z. Sun, X. Wang, and C. P. Sun, Phys. Rev. A \textbf{75},
                    062312 (2007).
\bibitem{lfw09pra}  Z.-G. Li, F.-S. Fei, Z. D. Wang and W.-M. Liu,
                    Phys. Rev. A \textbf{79}, 024303 (2009).
\bibitem{fan09pra}  Z. Liu and H. Fan, Phys. Rev. A \textbf{79}, 064305 (2009).
\bibitem{ajh10pra}  Q.-J. Tong, J.-H. An, H.-G. Luo, and C. H. Oh,
                    Phys. Rev. A \textbf{81}, 052330 2010.
\bibitem{ysw10pra}  Y. S. Weinstein, Phys. Rev. A \textbf{82}, 032326 (2010).
\bibitem{tyu09sci}  T. Yu and J. H. Eberly, Science \textbf{323}, 598 (2009).
\bibitem{alm07sci}  M. P. Almeida, F. de Melo, M. Hor-Meyll, A. Salles,
                    S. P. Walborn, P. H. Souto Ribeiro, L. Davidovich,
                    Science \textbf{316}, 579 (2007).
\bibitem{lau07prl}  J. Laurat, K. S. Choi, H. Deng, C. W. Chou, and H. J. Kimble,
                    Phys. Rev. Lett. \textbf{99}, 180504 (2007).
\bibitem{clo08prl}  C. E. L\'{o}pez, G. Romero, F. Lastra, E. Solano, and
                    J. C. Retamal, Phys. Rev. Lett. \textbf{101}, 080503 (2008).
\bibitem{byw09pra}  Y.-K. Bai, M.-Y. Ye, and Z. D. Wang,
                    Phys. Rev. A \textbf{80}, 044301 (2009).
\bibitem{nie00book} M. A. Nielsen and I. L. Chuang, \emph{Quantum Computation
                    and Quantum Information} (Cambridge University Press,
                    Cambridge, England, 2000), p20.
\bibitem{woo98prl}  W. K. Wootters, Phys. Rev. Lett. \textbf{80}, 2245 (1998).
\bibitem{expl2}     The square root of the eigenvalues of matrix $R_{c_1c_2}$ are
                    $\sqrt{\lambda_1}=\alpha\beta\xi^2+\sqrt{\beta^2\xi^4(\alpha^2 +
                    \beta^2cr^4\chi^4)}$, $\sqrt{\lambda_2}=-\alpha\beta\xi^2
                    +\sqrt{\beta^2\xi^4(\alpha^2 + \beta^2cr^4\chi^4)}$, and
                    $\sqrt{\lambda_3}=\sqrt{\lambda_4}=\beta^2\xi^2\chi^2cr^2$ with
                    $cr=\mbox{cos}(\gamma/2)$.
                    The square root of the eigenvalues of matrix
                    $R_{r_1r_2}$ are same to those of matrix
                    $R_{c_1c_2}$ after exchanging the parameters
                    $\xi$ and $\chi$.
\bibitem{car04prl}  A. R. R. Carvalho, F. Mintert, and A. Buchleitner,
                    Phys. Rev. Lett. \textbf{93}, 230501 (2004).
\bibitem{byw07pra}  Y.-K. Bai, D. Yang, and Z. D. Wang, Phys. Rev. A
                    \textbf{76}, 022336 (2007).
\bibitem{baw08pra}  Y.-K. Bai and Z. D. Wang, Phys. Rev. A \textbf{77},
                    032313 (2008).
\bibitem{ren08pra}  X.-J. Ren, W. Jiang, X. Zhou, Z. W. Zhou, and G. C. Guo,
                    Phys. Rev. A \textbf{78}, 012343 (2008).
\bibitem{won01pra}  A. Wong and N. Christensen, Phys. Rev. A \textbf{63},
                    044301 (2001).
\bibitem{ckw00pra}  V. Coffman, J. Kundu, and W. K. Wootters,
                    Phys. Rev. A \textbf{61}, 052306 (2000).
\bibitem{loh06prl}  R. Lohmayer, A. Osterloh, J. Siewert, and A. Uhlmann,
                    Phys. Rev. Lett.
                    \textbf{97}, 260502 (2006).
\bibitem{byw08pra}  Y.-K. Bai, M.-Y. Ye, and Z. D. Wang, Phys. Rev. A
                    \textbf{78}, 062325 (2008).
\bibitem{osb06prl}  T. J. Osborne and F. Verstraete, Phys. Rev. Lett.
                    \textbf{96}, 220503 (2006).
\bibitem{dlo98pra}  D. Loss and D. V. DiVincenzo, Phys. Rev. A \textbf{57},
                    120 (1998).
\bibitem{ssl01pns}  S.-S. Li, G.-L. Long, F.-S. Bai, S.-L. Feng, and H.-Z. Zheng,
                    Proc. Natl. Acad. Sci. U.S.A. \textbf{98}, 11847 (2001).
\bibitem{xgw06pra}  X. Wang and Z. D. Wang, Phys. Rev. A \textbf{73},
                    064302 (2006).
\bibitem{bel07prl}  B. Bellomo, R. Lo Franco, and G. Compagno, Phys. Rev. Lett.
                    \textbf{99}, 160502 (2007).
\bibitem{nov08pra}  E. Novais, E. R. Mucciolo, and H. U. Baranger,
                    Phys. Rev. A \textbf{78}, 012314 (2008).
\bibitem{kon08nat}  T. Konrad, F. De Melo, M Tiersch, C. Kasztelan, A. Aragao,
                    and A. Buchleitner,
                    Nature Physics \textbf{4}, 99 (2008).
\bibitem{wan10epr}  X.-B. Wang, Z.-W. Yu and J.-Z. Hu, arXiv:1001.0156.

\bibitem{expl1}     Due to the space of system $c_1r_1$ is
                    spanned by $\{\ket{00},\ket{01},\ket{10}\}$, we
                    have $(\hat{a}^{\dagger}\hat{a}-1)\hat{a}^{\dagger}\hat{a}
                    =0\Rightarrow \hat{N}^2=\hat{N}$. Similarly, according to the
                    space of system $c_1r_1$, we have $\hat{a}^{\dagger}
                    \hat{b}\hat{b}^{\dagger}\hat{b}\hat{N}=0$.
                    Combining it with the commutation relation of $\hat{b}$
                    and $\hat{b}^{\dagger}$, we can derive
                    $\hat{a}^{\dagger}\hat{N}\hat{b}_k=0$.

\end{thebibliography}
\end{document}